\documentclass[reprint,pre]{revtex4-1}
\usepackage{bm,epsfig,graphicx,color,amssymb,amsmath}

\begin{document}

\title{Exciton Relaxation in Carbon Nanotubes via Electronic-to-Vibrational Energy Transfer}

\author{Kirill A. Velizhanin}
\email{kirill@lanl.gov}
\affiliation{Theoretical Division, Los Alamos National Laboratory, Los Alamos, New Mexico 87545, USA}

\begin{abstract}
Covalent functionalization of semiconducting single-wall carbon nanotubes (CNT)
introduces new photoluminescent emitting states. Theses states are spatially
localized at around functionalization sites and strongly red-shifted
relative to the emission commonly observed from the nanotube band-edge
exciton state. A particularly important feature of these localized
exciton states is that, because the exciton is no longer free to diffusively
sample photoluminescent quenching sites along the CNT length, its lifetime is significantly extended. We have recently
demonstrated that an important relaxation channel of such localized
excitons is the electronic-to-vibrational energy transfer (EVET).
This process is analogous to the F\"orster resonance energy transfer
(FRET) except the final state of this process is not electronically,
but vibrationally excited molecules of the surrounding medium (e.g.,
solvent). In this work we develop the general theory of EVET, and
apply it to the specific case of EVET-mediated relaxation of defect-localized
excitons in covalently functionalized CNT. The resulting EVET relaxation
times are in good agreement with experimental data. 
\end{abstract}

\maketitle

\section{Introduction}

Covalent functionalization of semiconducting single-wall carbon nanotubes
(CNT) by oxygen \cite{Ghosh-2010-1656}, aryl \cite{Piao-2013-840},
and alkyl groups \cite{Kwon-2016-6878}, with the latter two classes
creating sp$^{3}$ defects, introduces a new photoluminescent (PL)
emitting state that is strongly red-shifted from the emission commonly
observed from the CNT band-edge exciton state \cite{OConnel-2002-593}.
In addition to being the source of new photophysical behaviors, these
states are drawing significant interest as the basis for potential
emerging functionality such as sensing and imaging \cite{Ghosh-2010-1656,Kwon-2015-3733,Shiraki-2016-12972},
photon upconversion \cite{Akizuki-2015-8920,Maeda-2016-16916}, and
room-temperature single photon emission \cite{Ma-2015-671,He-2017-577}.
Many of these functionalities arise due to localization of the diffusive
band-edge exciton at the defect site \cite{Miyauchi-2013-715,Hartmann-2016-8355,Ma-2014-10782},
which in turn leads to significant extension of the exciton lifetime
\cite{Ma-2015-671,He-2017-577,Miyauchi-2013-715,Hartmann-2016-8355}.
This is because the exciton, once trapped at the defect site, is no
longer free to diffusively sample PL quenching sites along the length
of the carbon nanotube \cite{Hertel-2010-7161}. Since the long exciton
lifetime (hundreds of picoseconds) becomes critical for proposed functionalities,
much effort has been directed lately towards understanding the nature
of relaxation channels of the trapped exciton in functionalized CNTs
\cite{Kim-2016-11268,Hartmann-2016-8355,Ma-2017-16143,He-2018-8060}.
In particular, we have recently suggested that resonance energy
transfer between the trapped exciton in CNT and solvent vibrational
modes can be an important exciton relaxation channel \cite{He-2018-8060}.

Electronic-to-vibrational energy transfer (EVET) process, where energy
is transferred from an exciton to solvent vibrational degrees of freedom,
has been previously shown to be an important exciton relaxation channel
in colloidal semiconductor nanoparticles with characteristic EVET
times on the order of a few hundred nanoseconds \cite{Aharoni-2008-057404,Wen-2016-4301}.
The prerequisite for EVET is existance of solvent vibrational modes
with phonon energies of $\sim1\,{\rm eV}$. No single fundamental
vibrational mode of this energy exists in solvent, but anharmonic effects give
rise to the so called overtone and combination modes \cite{Falk-1966-1699,Wen-2016-4301}.
For example, two representative combination modes in water are $2\nu_{1}+\nu_{3}\approx9870\,{\rm cm^{-1}\approx}1.22\,{\rm eV}$
and $\nu_{1}+\nu_{2}+\nu_{3}\approx8250\approx1.02\,{\rm eV}$, where
$\nu_{1}=3261\,{\rm cm^{-1}}$, $\nu_{2}=1639\,{\rm cm}^{-1}$ and
$\nu_{3}=3351\,{\rm cm^{-1}}$ are the symmetrical stretching, bending
and asymmetrical stretching fundamental vibrational modes, respectively,
for ${\rm H_{2}O}$ molecule. Interaction of these modes with external
electric field is fully encoded by complex dielectric function, or,
equivalently, by the refractive index and absorption coefficient \cite{Wen-2016-4301}.
For example, the existence of these modes is the reason water is not
completely transparent in the near-infrared spectral region, so that
the absorption coefficient is around $\sim1\,{\rm cm^{-1}}$ at $\lambda=1150-1300\,{\rm nm}$
\cite{He-2018-8060}.

In Ref. \cite{He-2018-8060} we suggested that relaxation times of
a localized exciton in CNTs, associated with EVET, can be as short
as $\sim100-200$ ps. This is significantly faster that that in semiconductor
nanoparticles due to much smaller characteristic distances between
exciton and absorbing medium for CNTs, $\lesssim 0.5-0.7$ nm \cite{He-2018-8060},
compared to that of $\approx3$ nm for semiconductor nanoparticles
\cite{Aharoni-2008-057404,Wen-2016-4301}. The main goal of the present
work is to develop a theory for EVET-mediated relaxation of a localized
exciton in CNTs. To this end, we first develop a general formalism
of EVET, Sec. \ref{sec:Gen_Theory}. This is done by deriving the
effective response function of the exciton-bearing system in the presence
of environment, and then extracting the exciton lifetime from the
self-energy of the response function. Then, we apply the developed
formalism to the cases of (i) simple 1D system near a flat dielectric
interface, Sec. \ref{sec:1D-flat}, and (ii) spherical semiconductor
nanoparticle, Sec. \ref{sec:EVET_QD}. The latter section, where we
re-derive the result previously obtained by Wen et al. \cite{Wen-2016-4301},
is needed as a sanity check of the general theory. Finally in Sec.
\ref{sec:EVET-CNT}, we obtain a simple closed-form expression for
EVET rates for the localized exciton in CNT and estimate the EVET
exciton relaxation times at realistic conditions. The resulting times
are $\sim100$ ps, which, considering the level of approximations,
is in good agreement with experimentally observed $\tau_{{\rm EVET}}\sim200$
ps in Ref. \cite{He-2018-8060}. Section \ref{sec:Conclusion} concludes.

\section{General Theory of Resonance Energy Transfer\label{sec:Gen_Theory}}

We split the entire space into two regions assuming that electronic/vibrational
wave functions are localized within them, so that the interaction
between these regions proceeds only through electromagnetic field. Specific
electronic transition $0\leftrightarrow1$ of energy $E_{10}=\hbar\omega_{10}$
in the first region is designated as the subsystem $I$. All the other (non-resonant)
electronic/vibrational transitions in the first region and all the electronic/vibrational transitions in the second region are designated as the subsystem $II$. The response function
$\chi$ of some generic system to the external electric potential
$V({\bf r},\omega)$ is introduced as
\begin{equation}
\rho({\bf r},\omega)=\int d{\bf r}'\,\chi({\bf r},{\bf r}';\omega)V({\bf r}',\omega),\label{eq:response}
\end{equation}
where $\rho({\bf r},\omega)$ is the induced charge density. Vector
variables are typed in bold. The response function for the isolated
subsystem $I$ in Lehmann representation is \cite{Giuliani-Vignale-2005-Liquid}
\begin{equation}
\chi_{I}({\bf r},{\bf r}';\omega)=\frac{1}{\hbar}\left[\frac{\rho_{01}({\bf r})\rho_{10}({\bf r}')}{\omega-\omega_{10}+i\delta}-\frac{\rho_{10}({\bf r})\rho_{01}({\bf r}')}{\omega+\omega_{10}+i\delta}\right],
\end{equation}
where $\rho_{10}({\bf r})=-e\langle1|\hat{\varphi}^{\dagger}({\bf r})\hat{\varphi}({\bf r})|0\rangle$
is the transition charge density and $\hat{\varphi}$ ($\hat{\varphi}^{\dagger}$)
is the electronic annihilation (creation) field operator. Spin indices
are omitted in this work. In what follows we assume a non-magnetic
system, so that localized wavefunctions can be chosen real, which
in turn yields $\rho_{01}({\bf r})=\rho_{10}({\bf r})$, and therefore
\begin{equation}
\chi_{I}({\bf r},{\bf r}';\omega)=\frac{1}{\hbar}\frac{2\omega_{10}}{\omega^{2}-\omega_{10}^{2}+2i\omega\delta}\rho_{10}({\bf r})\rho_{10}({\bf r}').\label{eq:chiI_real}
\end{equation}

\subsection{Effective response function}

Imagine that we are able to fully solve the Poisson problem for the
subsystem $II$ so the electrostatic potential at position ${\bf r}$
induced by a unit charge located at ${\bf r}'$, fully incorporating
the dielectric response of $II$, is given by Green's function $V_{II}({\bf r},{\bf r}';\omega)$.
Then, the fully self-consistent charge density induced within subsystem
$I$ as a response to external potential $V_{e}({\bf r};\omega)$ acting
only on subsystem $I$ is (symbolically) $\rho_{I}=\chi_{I}V_{tot}$,
where $V_{tot}=V_{e}+V_{II}\rho_{I}$, resulting in
\begin{equation}
\rho_{I}=\chi_{I}V_{e}+\chi_{I}V_{II}\rho_{I}.
\end{equation}
This equation can be solved for the induced charge density as $\rho_{I}=\chi_{I}^{(eff)}V_{e}$,
where
\begin{equation}
\chi_{I}^{(eff)}({\bf r},{\bf r}';\omega)=\frac{1}{\hbar}\frac{2\omega_{10}\rho_{10}({\bf r})\rho_{10}({\bf r}')}{\omega^{2}-\omega_{10}^{2}-2\omega_{10}\Sigma(\omega)/\hbar}
\end{equation}
is the effective response function of the subsystem $I$ in the presence
of subsystem $II$. Self-energy
\begin{equation}
\Sigma(\omega)=\int d{\bf r}d{\bf r}'\,\rho_{10}({\bf r})V_{II}({\bf r},{\bf r}';\omega)\rho_{10}({\bf r}')\label{eq:self_energy}
\end{equation}
comes from the interaction with subsystem $II$. The real part of
self-energy at $\omega=\omega_{0}$ constitutes the energy shift of
the $0\leftrightarrow1$ transition , whereas the imaginary part produces
the decay rate of exciton population due to Ohmic losses to the environment
(subsystem $II$)
\begin{align}
\Gamma_{I}&=-2\Sigma''(\omega_{10})/\hbar\nonumber\\
&=-\frac{2}{\hbar}\int d{\bf r}d{\bf r}'\,\rho_{10}({\bf r})V_{II}''({\bf r},{\bf r}';\omega_{10})\rho_{10}({\bf r}'),\label{eq:dec_rate}
\end{align}
where double prime stands for the imaginary part. 

\section{1D system near flat interface\label{sec:1D-flat}}

Within the envelope function approximation \cite{YuCardona-Fundamentals-1999,Haug-Koch-2009},
wavefunctions for single-particle electronic states in a semiconductor
nanostructure can be represented as $\Psi_{i}({\bf r})=\phi_{i}({\bf r})u_{[i]}({\bf r})$,
where $i$ is the state index, and $[i]=c,v$ for $i$ pointing to
a state within the conduction or valence band, respectively. Bloch
functions $u_{c,v}({\bf r})$ are periodic over the lattice, and envelope
functions $\phi_{i}({\bf r})$ vary slowly over the unit cell. Quantum
mechanical amplitude for $j\rightarrow i$ transition due to interaction
with external potential $V({\bf r})$ is given by $-e\int d{\bf r}\,\Psi_{i}^{*}({\bf r})V({\bf r})\Psi_{j}({\bf r})$.
Transition charge density is then naturally introduced as $\rho_{ij}({\bf r})=-e\Psi_{i}^{*}({\bf r})\Psi_{j}({\bf r})$.
In the presence of the electron-hole interaction, electron and hole
are being scattering within respective bands in the correlated manner,
and so the exciton transition density is

\begin{equation}
\rho({\bf r})=\sum_{i\in c}\sum_{j\in v}B_{ij}\rho_{ij}({\bf r})=-e\sum_{i\in c}\sum_{j\in v}B_{ij}\Psi_{i}^{*}({\bf r})\Psi_{j}({\bf r}),\label{eq:tr_charge_dens}
\end{equation}
where $B_{ij}$ are the coefficients of expansion of the total exciton
wavefunction into the single-particle (electron and hole) ones. These
coefficients can generally be found by solving Bethe-Salpeter equation
\cite{Bechstedt-2015-Many-Body-Approach}. Explicitly separating
slowly and rapidly changing terms, Eq. (\ref{eq:tr_charge_dens})
can be rewritten as
\begin{equation}
\rho({\bf r})=\Phi({\bf r})\rho_{uc}({\bf r})
\end{equation}
where the slowly varying exciton envelope wavefunction is normalized
as $\int d{\bf r}\,\Phi^{2}({\bf r})=1$, and the lattice-periodic
part is $\rho_{uc}({\bf r})\propto u_{c}^{*}({\bf r})u_{v}({\bf r}$).
Importantly, integral of $\rho_{uc}$ over the unit cell volume vanishes
exactly due to orthogonality of $u_{c}({\bf r})$ and $u_{v}({\bf r})$.
However, dipole moment of the unit cell, $\int d{\bf r}\,{\bf r}\rho_{uc}({\bf r})$,
does not generally vanish. Exciton transition charge density can thus
be thought of as comprising many small transition dipoles, one for
each unit cell of the semiconductor. 

\begin{figure}

\includegraphics[width=3.2in]{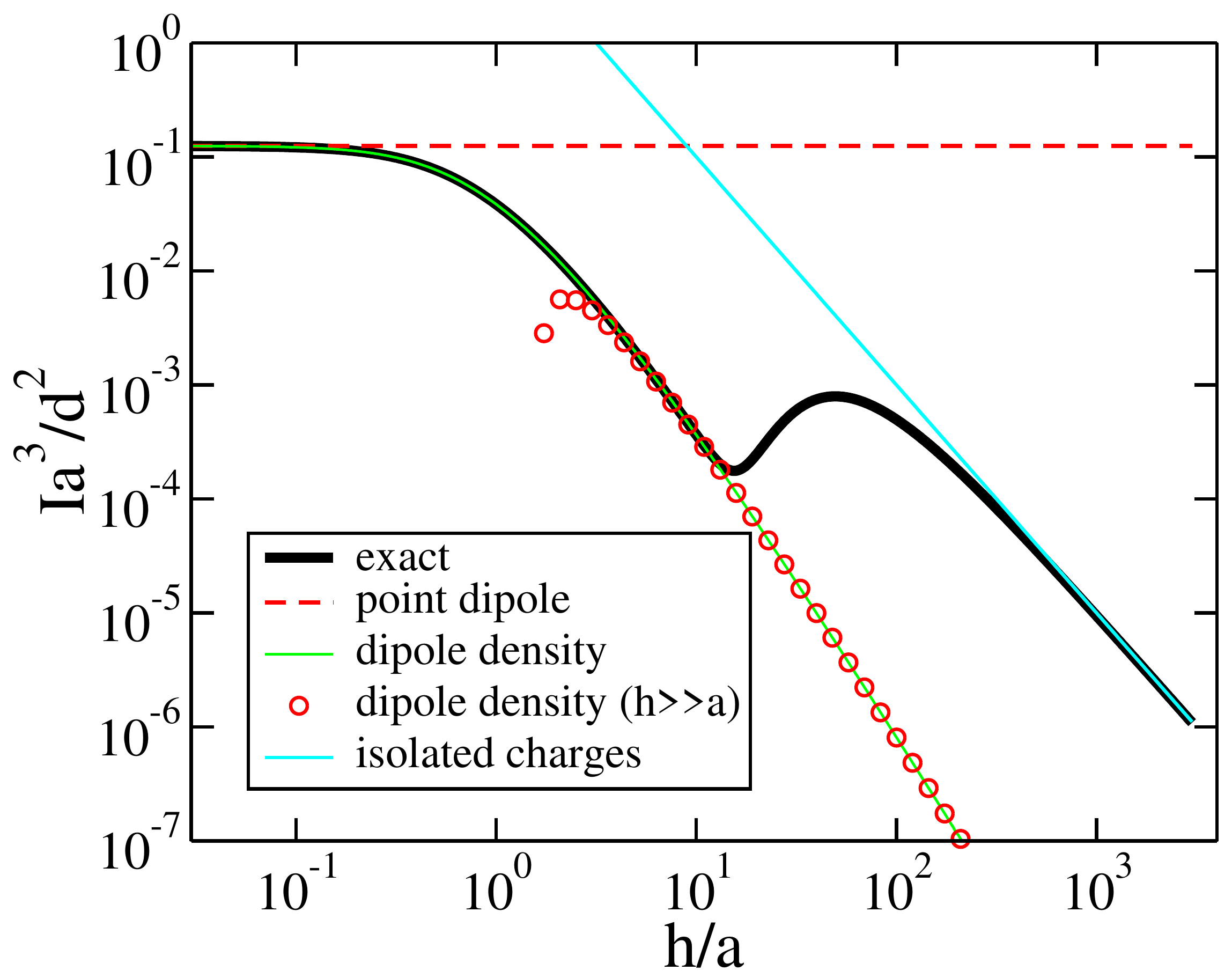}\caption{\label{fig:1d_flat}Equation (\ref{eq:I_1D}), normalized by $d^{2}/a^{3}$,
evaluated directly (thick black line) and using various approximation
(colored lines and circles) for the Gaussian envelope function. Results
for the exponential envelope function are very similar. We set $b=0.1h$.}
\end{figure}

To develop a general intuition for how the EVET decay rate depends on various
geometric parameters of a system, we consider a flat dielectric interface and
a simple 1D system whose transition density can be represented by
a collection of point charges. The distance between the 1D system
and the dielectric interface is $a$ and the transition density of
the former is given by
\begin{equation}
\rho(x)=\Phi(x)\rho_{uc}(x),\label{eq:td_1D}
\end{equation}
where $\rho_{uc}(x)=f\sum_{n}\left[\delta(x-b/2-bn)-\delta(x-bn)\right]$
encodes the transition dipole of the unit cell via two point charges.
Prefactor $f$ is obtained from setting the total transition dipole
moment of the system to $\int dx\,x\rho(x)=d$. Unit cell length is $b$.
EVET decay rate near the flat dielectric interface can be
evaluated by first finding the Green's function $V_{II}({\bf r},{\bf r}';\omega)$ with the help of the method of image charges, and then substituting this Green's function into Eq. (\ref{eq:dec_rate}). Accordingly, the dependence
of the decay rate on geometry of the system is fully encoded by the
following expression
\begin{equation}
I=\sum_{i,j}\frac{Q_{i}Q_{j}}{\sqrt{(x_{i}-x_{j})^{2}+(2a)^{2}}},\label{eq:I_1D}
\end{equation}
where $Q_{i}$ are the point charges constituting the transition density
(\ref{eq:td_1D}), and $x_{i}$ are their positions. In what follows
we consider various limiting cases where Eq. (\ref{eq:I_1D}) can
be evaluated analytically or almost analytically. To this end, we
consider two possibilities for the exciton envelope function: exponential
and Gaussian. Exponential envelope function is given by
\begin{equation}
\Phi(x)=h^{-1/2}e^{-|x|/h},\label{eq:exp_envelope}
\end{equation}
where $h$ is the characteristic width of the envelope.
Its Fourier transform is
\begin{equation}
\Phi_{k}=\int_{-\infty}^{\infty}dx\,\Phi(x)e^{-ikx}=\frac{2h^{1/2}}{1+k^{2}h^{2}}.\label{eq:exp_env_Fourier}
\end{equation}
The second example is the Gaussian envelope function
\begin{equation}
\Phi(x)=\left(2h\right)^{-1/2}e^{-\frac{\pi x^{2}}{8h^{2}}}\label{eq:gauss_envelope}
\end{equation}
with the Fourier transform
\begin{equation}
\Phi_{k}=\int_{-\infty}^{\infty}dx\,\Phi(x)e^{-iqx}=2h^{1/2}e^{-\frac{2q^{2}h^{2}}{\pi}}.\label{eq:gauss_env_Fourier}
\end{equation}
The numerical prefactor of $\pi/8$ in the exponent of Eq. (\ref{eq:gauss_envelope})
is chosen intentionally so that $\int dx\,\Phi(k)=\Phi_{k=0}=2h^{1/2}$
is the same for the both envelope functions. Black line in Fig. \ref{fig:1d_flat}
shows the result of direct numerical evaluation of Eq. (\ref{eq:I_1D})
for the Gaussian envelope function.
We set $b=0.1h$ so that there is many units cells within the envelope,
otherwise envelope function approximation would break down \cite{YuCardona-Fundamentals-1999,Haug-Koch-2009}.

\subsection{$h\gg b\gg a$}

In this case, $b\gg a$ and so each transition charge does not interact
with images of other charges. Summation in Eq. (\ref{eq:I_1D}) is
then dominated by the diagonal ($i=j$) terms
\begin{equation}
I=\frac{1}{2a}\sum_{i}Q_{i}^{2}.
\end{equation}
Since $b=0.1h\ll h$ and, therefore, there is a lot of unit cells
within the envelope, summation can be substituted with integration
as
\begin{equation}
I=\frac{4d^{2}}{ab}\frac{\int dx\,\Phi^{2}(x)}{\left(\int dx\,\Phi(x)\right)^{2}}=\frac{d^{2}}{abh}.
\end{equation}
This expression is plotted as a cyan line in Fig. \ref{fig:1d_flat}
and is seen to agree well with exact numerical result (black line)
at $h/a\gtrsim 200$, or equivalently $b/a\gtrsim 20$. This rather
slow convergence to exact results is due to the long-range nature
of Coulomb interaction.

\subsection{$b\ll a\ll h$}

A single transition charge interacts with a large number of image
charges at $b\ll a$, and so the summations in Eq. (\ref{eq:I_1D})
can be substituted with integrations
\begin{align}
I&=\int dx\int dx'\,\frac{\rho(x)\rho(x')}{\sqrt{(x-x')^{2}+(2a)^{2}}}\nonumber\\
&=\frac{2}{\pi}\int_{0}^{\infty}dk\,\rho_{-k}\rho_{k}K_{0}(2ka),
\end{align}
where we performed Fourier transformation of the kernel, which yielded
modified Bessel function of the second kind $K_{0}$. Dipole moment
(or polarization) density $P(x)$ is defined via $\partial_{x}P=-\rho$,
or $\rho_{k}=-ikP_{k}$, and so
\begin{equation}
I=\frac{2}{\pi}\int_{0}^{\infty}dk\,k^{2}P_{k}P_{-k}K_{0}(2ka).\label{eq:I_Pk}
\end{equation}
Introduction of the dipole moment density is convenient because, first,
using definition (\ref{eq:td_1D}) it is a simple matter to show that
$P_{k}$ is exactly proportional to $\Psi_{k}$ if the momentum is
not too large ($kb\ll1$). Secondly, the definition of $P$ implies
that $P_{k=0}$ equals the total transition dipole moment $d$. This
means that we can directly write $P_{k}=\frac{d}{1+k^{2}h^{2}}$ from
Eq. (\ref{eq:exp_env_Fourier}) for the exponential envelope function,
resulting in
\begin{equation}
I=\frac{2d^{2}}{\pi}\int_{0}^{\infty}dk\,k^{2}\frac{K_{0}(2ka)}{(1+k^{2}h^{2})^{2}}.\label{eq:I_Psmooth_exp}
\end{equation}
For the Gaussian envelope function this equation becomes
\begin{equation}
I=\frac{2d^{2}}{\pi}\int_{0}^{\infty}dk\,k^{2}e^{-4k^{2}h^{2}/\pi}K_{0}(2ka).\label{eq:I_Psmooth_gauss}
\end{equation}
This last equation is numerically integrated, yielding thin green
line in Fig. \ref{fig:1d_flat}. As expected, this line agrees with the exact numerical result, when $b$ is smaller than a certain threshold. More specifically,
the agreement is already excellent when $b/a\lesssim1.3$ This is
a very important result since it implies that the specific distribution
of the transition density within the unit cell is irrelevant, and the continuous
and smooth $P(x)\propto\Phi(x)$ can be used when $b\lesssim a$.

Integral in Eq. (\ref{eq:I_Psmooth_exp}) can be evaluated analytically
if we assume $h\gg a$, which allows one to use the small-argument
asymptotic form of $K_{0}$ \cite{Abramowitz-Stegun-1965}
\begin{align}
I&=-\frac{2d^{2}}{\pi}\int_{0}^{\infty}dk\,\frac{k^{2}\left(\ln ka+\gamma\right)}{(1+k^{2}h^{2})^{2}}\nonumber\\
&=\frac{d^{2}}{2h^{3}}\left(\ln\frac{h}{a}-\gamma-1\right),
\end{align}
where $\gamma=0.5772...$ is Euler--Mascheroni constant.

For the gaussian envelope, Eq. (\ref{eq:gauss_env_Fourier}), one
has $P_{k}=de^{-2k^{2}h^{2}/\pi}$, and therefore
\begin{align}
I&=-\frac{2d^{2}}{\pi}\int_{0}^{\infty}dk\,k^{2}e^{-4k^{2}h^{2}/\pi}\left(\ln ka+\gamma\right)\nonumber\\
&=\frac{\pi d^{2}}{32h^{3}}\left(\ln\frac{16h^{2}}{\pi a^{2}}-\gamma-2\right).
\end{align}
This last expression is plotted with red circles in Fig. \ref{fig:1d_flat}.
It is seen to agree with (i) Eq. (\ref{eq:I_Psmooth_gauss}) at $h/a\gtrsim 4$
and (ii) exact result (black line) in a window between $h/a\gtrsim 4$
and $h/a\lesssim13$. This last upper boundary follows of course from
$b/a\lesssim1.3$ condition of applicability of Eq. (\ref{eq:I_Psmooth_gauss}).

\subsection{$b,h\ll a$}

In this case we effectively have a point dipole interacting with the
dielectric interface. Method of image charges is applied and the resulting
dipole-dipole interaction energy is \cite{Jackson-1998}
\begin{equation}
I=\frac{d^{2}}{(2a)^{3}},
\end{equation}
which is plotted as a dashed red horizontal line in Fig. \ref{fig:1d_flat} and seed to agree with the exact numerical results at $h/a \lesssim 0.3$.

In the context of above derivations, this expression can also
be obtained from Eq. (\ref{eq:I_Pk}) by assuming strongly localized
$P(x)$. This results in $k$-independent $P_{k}\approx d$, leading
to the same result
\begin{equation}
I=\frac{2}{\pi}d^{2}\int_{0}^{\infty}dk\,k^{2}K_{0}(2ka)=\frac{d^{2}}{(2a)^{3}}.
\end{equation}

\section{EVET for a quantum dot\label{sec:EVET_QD}}

In this section we consider the EVET decay rate for a realistic system
- spherical semiconductor nanoparticle (quantum dot) of radius $a$
in a solvent. Electrodynamically, it is equivalent to a spherical
cavity of radius $a$ with dielectric constants $\epsilon_{in}$ and
$\epsilon_{out}$ inside (nanoparticle) and outside (solvent) the
cavity, respectively. We have a non-vanishing transition transition
charge density $\rho({\bf r})$ only within the cavity and we want
to find the corresponding exciton decay rate. The Poisson equation
inside is
\begin{equation}
-\Delta\varphi=\frac{4\pi}{\epsilon_{in}}\rho({\bf r}).
\end{equation}
Outside, it is the homogenous Poisson (i.e., Laplace) equation, whose
solution must be matched to the one inside using the appropriate boundary
conditions. To this end, we first expand the transition density into
spherical harmonics as
\begin{equation}
\rho({\bf r})=\sum_{l,m}\rho_{lm}(r)Y_{l}^{m}({\bf r}).\label{eq:rho_sphere_exp}
\end{equation}
Second, we initially assume that $\rho_{lm}(r)=\sigma\delta(r-R)$
where $R<a$. Homogeneous Poisson equation for given spherical symmetry
$Y_{l}^{m}$ is
\begin{equation}
\left[-\partial_{r}^{2}-\frac{2}{r}\partial_{r}+\frac{l(l+1)}{r^{2}}\right]\varphi(r)=0.
\end{equation}
with solutions $\varphi_{1}=r^{l}$ and $\varphi_{2}=r^{-(l+1)}$.
We solve a radial boundary problem with three regions: (I) $r\in(0,R)$,
(II) $r\in(R,a)$, (III) $r\in(a,\infty)$. The potential in these
three regions are (assuming non-singular behavior at $r\rightarrow0$
and vanishing potential at $r\rightarrow\infty$)
\begin{align}
\varphi_{I} & =A\varphi_{1}(r),\\
\varphi_{II} & =B\varphi_{1}(r)+C\varphi_{2}(r),\nonumber \\
\varphi_{III} & =D\varphi_{2}(k),\nonumber 
\end{align}
where coefficients $A,B,C,D$ are found from the boundary conditions:
\begin{gather}
A\varphi_{1}(R)=B\varphi_{1}(R)+C\varphi_{2}(R),\\
B\varphi_{1}(a)+C\varphi_{2}(a)=D\varphi_{2}(a),\nonumber \\
\left.\frac{d}{dr}\left(B\varphi_{1}(r)+C\varphi_{2}(r)\right)\right|_{r=R}-\left.\frac{d}{dr}A\varphi_{1}(r)\right|_{r=R}=-4\pi\sigma/\epsilon_{in},\nonumber \\
\epsilon_{in}\left.\frac{d}{dr}\left(B\varphi_{1}(r)+C\varphi_{2}(r)\right)\right|_{r=a}=\epsilon_{out}\left.\frac{d}{dr}D\varphi_{2}(r)\right|_{r=a},\nonumber 
\end{gather}
where the first two equations match the electrostatic potential at
the two boundaries, the third equation encodes the jump of the electric
field due to the surface charge at $r=R$, and the last one encodes
the jump of the electric field due to the the mismatch of the electric
fields due to the induced surface charge on the cavity surface. The
resulting coefficients are
\begin{align}
A & =\frac{4\pi\sigma R^{1-l}}{\epsilon_{in}(2l+1)}\left(1+\frac{(1+l)(\epsilon_{in}-\epsilon_{out})}{\epsilon_{out}+l(\epsilon_{in}+\epsilon_{out})}\left(\frac{R}{a}\right)^{2l+1}\right),\\
B & =\frac{4\pi\sigma}{\epsilon_{in}(2l+1)}\frac{(1+l)(\epsilon_{in}-\epsilon_{out})}{\epsilon_{out}+l(\epsilon_{in}+\epsilon_{out})}\frac{R^{l+2}}{a^{2l+1}},\nonumber \\
C & =\frac{4\pi\sigma}{\epsilon_{in}(2l+1)}R^{l+2},\nonumber \\
D & =\frac{4\pi\sigma}{\epsilon_{out}+l(\epsilon_{in}+\epsilon_{out})}R^{l+2}.\nonumber 
\end{align}
These coefficients encode both ``direct'' and ``induced'' fields.
To get the induced fields for e.g., coefficient $A$ we have to evaluate
$A^{(ind)}=A-A(\epsilon_{out}\rightarrow\epsilon_{in})$. The result
is (for fields inside the cavity)
\begin{align}
A^{(ind)} & =\frac{4\pi\sigma}{\epsilon_{in}(2l+1)}\frac{(1+l)(\epsilon_{in}-\epsilon_{out})}{\epsilon_{out}+l(\epsilon_{in}+\epsilon_{out})}\frac{R^{l+2}}{a^{2l+1}},\\
B^{(ind)} & =\frac{4\pi\sigma}{\epsilon_{in}(2l+1)}\frac{(1+l)(\epsilon_{in}-\epsilon_{out})}{\epsilon_{out}+l(\epsilon_{in}+\epsilon_{out})}\frac{R^{l+2}}{a^{2l+1}},\nonumber \\
C^{(ind)} & =0.\nonumber 
\end{align}
As expected $\varphi_{I}^{(ind)}$ matches smoothly into $\varphi_{II}^{(ind)}$
since induced potential is created by polarization of the interface
at $r=a$ and, therefore, do not have a singularity at $r=R$. The
resulting induced potential inside the cavity is
\begin{equation}
\varphi^{(ind)}(r)=\frac{4\pi\sigma}{\epsilon_{in}(2l+1)}\frac{(1+l)(\epsilon_{in}-\epsilon_{out})}{\epsilon_{out}+l(\epsilon_{in}+\epsilon_{out})}\frac{R^{l+2}}{a^{2l+1}}r^{l}.
\end{equation}
This is the potential from the $r=R$ shell. The full potential is
\begin{align}
\varphi_{lm}^{(ind)}(r)&=\frac{4\pi}{\epsilon_{in}(2l+1)}\frac{(1+l)(\epsilon_{in}-\epsilon_{out})}{\epsilon_{out}+l(\epsilon_{in}+\epsilon_{out})}\nonumber\\
&\times\frac{r^{l}}{a^{2l+1}}\int_{0}^{a}dR\,R^{l+2}\rho_{lm}(R).
\end{align}
The self-energy, Eq. (\ref{eq:self_energy}), for this transition density would be
\begin{align}
\Sigma&=\int_{{\rm cavity}}d{\bf r}\,\rho({\bf r})\varphi^{(ind)}({\bf r})\nonumber\\
&=\sum_{l,m}\int_{0}^{a}r^{2}dr\,\rho_{l,-m}(r)\varphi_{lm}^{(ind)}(r).\label{eq:self_en_spher}
\end{align}

We now need to discuss how to approximate $\rho({\bf r})$. By performing
Fourier transform of transition charge density (\ref{eq:tr_charge_dens})
and discarding large-momentum components corresponding to the internal
structure of $u_{c,v}({\bf r})$, one can show that the vector field
of transition dipole moment density can be written as ${\bf P}({\bf r})=P({\bf r}){\bf e}_{uc}$.
Here, $P({\bf r})\propto\Psi({\bf r})$ - exciton envelope wavefunction,
defined in Sec. \ref{sec:1D-flat}, and ${\bf e}_{uc}$ is a unit
vector in the direction of the transition dipole of the unit cell,
$\int_{uc}d{\bf r}\,{\bf r}u_{c}^{*}({\bf r})u_{v}({\bf r})$. Neglecting
the large-momentum components was justified in Sec. \ref{sec:1D-flat}
for cases where characteristic distance from the unit cell to the
dielectric interface is larger than the unit cell itself. This is
true for typical semiconductor quantum dots since their size, and,
therefore, typical distance between a unit cell and the dielectric
interface, is much larger than the lattice constant \cite{Norris-2010-Nanocrystals}.
The lowest-energy exciton transition in quantum dots with strong confinement
reduces to a product of single-particle electron and hole wavefunctions,
whose respective single-particle envelope functions are spherically
symmetric \cite{Norris-2010-Nanocrystals}. This results in spherically
symmetric exciton envelope function $\Phi({\bf r})=\Phi(r)$, and
therefore we have
\begin{equation}
{\bf P}({\bf r})=P(r){\bf e}_{uc}.\label{eq:Pr_sphere}
\end{equation}
Transition charge density is found from transition dipole moment density
as $\rho=-\nabla\cdot{\bf P}$. Orienting ${\bf e}_{uc}$ along $z$-axis
in Eq. (\ref{eq:Pr_sphere}) we find that the expansion of $\rho({\bf r})$
into spherical harmonics, Eq. (\ref{eq:rho_sphere_exp}), consists
of only a single term of $Y_{1}^{0}$ symmetry. Eq. (\ref{eq:self_en_spher})
then becomes
\begin{equation}
\Sigma=\left[\int_{0}^{a}r^{3}dr\,\rho_{1,0}(r)\right]^{2}\frac{8\pi}{3}\frac{\epsilon_{in}-\epsilon_{out}}{\epsilon_{in}+2\epsilon_{out}}\frac{1}{\epsilon_{in}a^{3}}.
\end{equation}
The resulting integral in square brackets is exactly proportional
to the transition dipole moment $\int_{0}^{a}r^{3}dr\,\rho_{1,0}(r)=\left(\frac{3}{4\pi}\right)^{1/2}d$,
so the result for the self-energy is
\begin{equation}
\Sigma=\frac{2(\epsilon_{in}-\epsilon_{out})}{\epsilon_{in}+2\epsilon_{out}}\frac{d^{2}}{\epsilon_{in}a^{3}}.\label{eq:self_energy_spherical}
\end{equation}
We further assume that $\epsilon_{in}$ is real and $\epsilon_{out}=\epsilon_{out}'+i\epsilon_{out}''$, where the imaginary part $\epsilon''_{out}$ is small. The resulting
decay rate is
\begin{equation}
\Gamma=-2\hbar^{-1}\Sigma''=\hbar^{-1}\frac{12\epsilon_{out}''}{\left[\epsilon_{in}+2\epsilon_{out}'\right]^{2}}\frac{d^{2}}{a^{3}},\label{eq:decay_spherical}
\end{equation}
where the dielectric constants are to be evaluated at the frequency corresponding
to the exciton transition energy. The resulting equation for $\Gamma$
is exactly the same as Eq. (16) in Ref. \cite{Wen-2016-4301}. The
most important observation here is that the substitution of the entire
transition charge density distribution with its dipole moment is not
an approximation - specific radial distribution of dipole moment density is irrelevant as long as ${\bf P}({\bf r})=P(r){\bf e}_{uc}$.
Eq. (\ref{eq:decay_spherical}) is thus the same for the both truly point transition dipole located at the center of the spherical cavity, or for the distributed transition dipole of e.g., an electronic transition of a semiconductor quantum dot.

\section{EVET for Carbon Nanotube\label{sec:EVET-CNT}}

Electrodynamically, an exciton transition in a CNT sitting inside solvent can be treated as follows. There
is a cylindrical cavity of radius $a$, dielectric constant inside
and outside are $\epsilon_{in}$ and $\epsilon_{out}$, respectively.
Radius $a$ is the CNT solvation radius. The transition charge density
$\rho(x,\xi)$, technically surface density, resides on a CNT of radius
$R<a$ inside the cavity, where $\xi=R\theta$ is the circumferential
coordinate. To facilitate the derivation, the charge density is assumed
axially periodic with period $L_{x}$, and therefore can be represented
by a Fourier series
\begin{equation}
\rho_{k,m}=\int_{0}^{L_{x}}dx\int_{0}^{L_{\xi}}d\xi\,\rho(x,\xi)e^{-ikx}e^{-ik_{\xi}\xi},
\end{equation}
where $L_{\xi}=2\pi R$, $k_{\xi}=m/R$ and $k=\frac{2\pi}{L_{x}}n$
so that the inverse transformation looks like
\begin{equation}
\rho(x,\xi)=\frac{1}{L_{x}}\frac{1}{L_{\xi}}\sum_{k}\sum_{m}\rho_{k,m}e^{ikx}e^{im\theta}.
\end{equation}
The final result will not depend on $L_{x}$ when it is large. We
now want to find an electric potential induced by charge density of
specific symmetry $\rho(x,\xi)=\frac{1}{L_{x}L_{\xi}}\rho_{k,m}e^{ikx}e^{ik_{\xi}\xi}$.
The homogeneous solution of the Poisson equation within the cylindrical
cavity, $\Delta\varphi=0$, is found via expanding into cylindrical
harmonics
\begin{equation}
\left(\partial_{r}^{2}+\frac{1}{r}\partial_{r}-k^{2}-\frac{m^{2}}{r^{2}}\right)\varphi=0.
\end{equation}
The solutions of this equation are modified Bessel functions of the
first, $I_{m}(kr)$, and second, $K_{m}(kr)$, kind, where $k$ and
$m$ are understood in the absolute sense, i.e., $k=|k|$ and $m=|m|$.
Now we solve a radial boundary problem with three regions: (I) $r\in(0,R)$,
(II) $r\in(R,a)$, (III) $r\in(a,\infty)$. The potential in these
three regions are (assuming non-singular behavior at $r\rightarrow0$
and vanishing potential at $r\rightarrow\infty$)
\begin{align}
\varphi_{I} & =AI_{m}(kr),\\
\varphi_{II} & =BI_{m}(kr)+CK_{m}(kr),\nonumber \\
\varphi_{III} & =DK_{m}(kr),\nonumber 
\end{align}
where coefficients $A,B,C,D$ are found from the boundary conditions:
\begin{gather}
AI_{m}(kR)=BI_{m}(kR)+CK_{m}(kR),\\
BI_{m}(ka)+CK_{m}(ka)=DK_{m}(ka),\nonumber \\
\left.\frac{d}{dr}\left(BI_{m}(kr)+CK_{m}(kr)\right)\right|_{r=R}-\left.\frac{d}{dr}AI_{m}(kr)\right|_{r=R}\nonumber\\
=-4\pi\frac{1}{L_{x}L_{\xi}}\rho_{k,m}/\epsilon_{in},\nonumber \\
\epsilon_{in}\left.\frac{d}{dr}\left(BI_{m}(kr)+CK_{m}(kr)\right)\right|_{r=a}=\epsilon_{out}\left.\frac{d}{dr}DK_{m}(kr)\right|_{r=a},\nonumber 
\end{gather}
where the first two equations match the electrostatic potential at
the two boundaries, the third equation encodes the jump of the electric
field due to the surface charge at $r=R$, and the last one encodes
the jump of the electric field due to the the mismatch of the electric
fields due to the induced surface charge on the cavity surface. The
exact solution is
\begin{align}
C_{m}(k) & =-\frac{1}{L_{x}L_{\xi}}\frac{4\pi\rho_{k,m}}{\epsilon_{in}k}\frac{I_{m}(kR)}{\mathcal{R}_{m}},\\
B_{m}(k) & =-\frac{1}{L_{x}L_{\xi}}\frac{4\pi\rho_{k,m}}{\epsilon_{in}k}\frac{(\epsilon_{out}-\epsilon_{in})I_{m}(kR)K_{m}(ka)K_{m}'(ka)}{\mathcal{R}_{m}\mathcal{A}_{m}},\nonumber \\
A_{m}(k) & =B_{m}(k)+\frac{K_{m}(kR)}{I_{m}(kR)}C_{m}(k)\nonumber \\
D_{m}(k) & =\frac{I_{m}(ka)}{K_{m}(ka)}B_{m}(k)+C_{m}(k),\nonumber 
\end{align}
where
\begin{align}
\mathcal{R}_{m} & =I_{m}(kR)K_{m}'(kR)-K_{m}(kR)I_{m}'(kR)=-\frac{1}{kR},\\
\mathcal{A}_{m} & =\epsilon_{in}K_{m}(ka)I_{m}'(ka)-\epsilon_{out}I_{m}(ka)K_{m}'(ka)\nonumber\\
&=\frac{\epsilon_{in}}{ka}-(\epsilon_{out}-\epsilon_{in})I_{m}(ka)K_{m}'(ka),\nonumber 
\end{align}
where Abel's identity for modified Bessel functions was used. 

The expressions for $A_{m}(k)$, $B_{m}(k)$, $C_{m}(k)$ and $D_{m}(k)$
have contributions from both ``direct'' and induced potentials.
Direct contribution can be found by equating $\epsilon_{out}$ to $\epsilon_{in}$.
Induced potential is then found by subtraction. The induced potential
at $r=R$ is
\begin{gather}
\varphi^{(ind)}(x,\xi)=B_{m}(k)I_{m}(kR)e^{ikx}e^{ik_{\xi}\xi}\nonumber\\
=-\frac{1}{L_{x}L_{\xi}}\frac{4\pi\rho_{k,m}}{\epsilon_{in}k}\frac{(\epsilon_{out}-\epsilon_{in})I_{m}^{2}(kR)K_{m}(ka)K_{m}'(ka)}{\mathcal{R}_{m}\mathcal{A}_{m}}e^{ikx}e^{ik_{\xi}\xi}.
\end{gather}
The resulting Fourier transform is
\begin{equation}
\varphi_{k,m}^{(ind)}=-\frac{4\pi\rho_{k,m}}{\epsilon_{in}k}\frac{(\epsilon_{out}-\epsilon_{in})I_{m}^{2}(kR)K_{m}(ka)K_{m}'(ka)}{\mathcal{R}_{m}\mathcal{A}_{m}}.\label{eq:phi_ind}
\end{equation}
The interaction of the charge density with the potential induced by
it reads as, Eq. (\ref{eq:self_energy}) 
\begin{align}
\Sigma&=\int_{0}^{L_{x}}dx\int_{0}^{L_{\xi}}d\xi\,\rho(x,\xi)\varphi^{(ind)}(x,\xi)\nonumber\\
&=\int_{0}^{L_{x}}dx\int_{0}^{L_{\xi}}d\xi\,\frac{1}{L_{x}L_{\xi}}\sum_{k,m}\rho_{-k,-m}e^{-ikx}e^{-ik_{\xi}\xi}\varphi^{(ind)}(x,\xi)\nonumber\\
&=\frac{1}{L_{x}L_{\xi}}\sum_{k,m}\sigma_{-k,-m}\varphi_{k,m}^{(ind)}.
\end{align}
Using Eq. (\ref{eq:phi_ind}) we have
\begin{align}
\Sigma&=\frac{1}{L_{x}L_{\xi}}\sum_{k,m}\rho_{-k,-m}\varphi_{k,m}^{(ind)}\nonumber\\
&=-\frac{1}{L_{x}L_{\xi}}\frac{4\pi(\epsilon_{out}-\epsilon_{in})}{\epsilon_{in}}\nonumber\\
&\times\sum_{k,m}\frac{\rho_{k,m}\rho_{-k,-m}}{k}\frac{I_{m}^{2}(kR)K_{m}(ka)K_{m}'(ka)}{\mathcal{R}_{m}\mathcal{A}_{m}}.\label{eq:self-int}
\end{align}
Slightly rewriting this expression for clarity and better readability
\begin{equation}
\Sigma=\frac{1}{2\pi L_{x}}\sum_{k,m}\rho_{k,m}\rho_{-k,-m}G_{m}(kR,ka),\label{eq:self_energy_fin}
\end{equation}
where
\begin{equation}
G_{m}(x,y)=\frac{4\pi(\epsilon_{out}-\epsilon_{in})}{\epsilon_{in}}\frac{I_{m}^{2}(x)K_{m}(y)K_{m}'(y)}{\left[\frac{\epsilon_{in}}{y}-(\epsilon_{out}-\epsilon_{in})I_{m}(y)K_{m}'(y)\right]}.\label{eq:kernel}
\end{equation}

Somewhat similarly to the case of quantum dot above, the transition
dipole moment density for a localized exciton in a CNT can be written
as
\begin{equation}
{\bf P}(x,\xi)=f\Phi(x){\bf e}_{x},\label{eq:P_Phi_ex}
\end{equation}
where $\Phi(x)$ is the envelope wavefunction of the localized exciton,
and $f$ is the normalizing factor that guarantees that $\int dxd\xi\,{\bf P}(x,\xi)=d{\bf e}_{x}$,
where $d$ is the magnitude of total transition dipole moment of the
exciton. Detailed derivation is present in Appendix \ref{sec:CNT_trans_dens}.
Transition charge density is
\begin{equation}
\rho(x,\xi)=-\nabla\cdot{\bf P}(x,\xi)=-fL_{x}^{-1/2}\sum_{k}ikB_{k}e^{ikx},
\end{equation}
where
\begin{equation}
B_{k}=L_{x}^{-1/2}\int_{0}^{L_{x}}dx\,\Phi(x)e^{-ikx}\label{eq:Psi_Fourier_ser}
\end{equation}
are the Fourier series coefficients for $\Phi(x)$. Accordingly,
\begin{align}
\rho_{k,m}&=\int_{0}^{L_{x}}dx\int_{0}^{L_{\xi}}d\xi\,\rho(x,\xi)e^{-iqx}e^{-ik_{\xi}\xi}\nonumber\\
&=-ifL_{x}^{1/2}L_{\xi}kB_{k}\delta_{m,0}.
\end{align}
Substituting this into expression for the self-energy becomes
\begin{equation}
\Sigma=\frac{f^{2}L_{\xi}^{2}}{2\pi}\sum_{q}k^{2}B_{k}B_{-k}G_{0}(kR,ka),
\end{equation}

\subsection{Delta-functional trap}

An important example is a wavefunction for a particle trapped in a
delta-function well, Eq (\ref{eq:exp_envelope}). The normalization
factor is $f=\frac{d}{2h^{1/2}L_{\xi}}$. The resulting self-energy
is
\begin{equation}
\Sigma=\frac{d^{2}}{8\pi h}\sum_{k}k^{2}B_{k}B_{-k}G_{0}(kR,ka).\label{eq:self_Bk}
\end{equation}
Comparing Eqs. (\ref{eq:Psi_Fourier_ser}) and (\ref{eq:exp_env_Fourier})
one has $B_{k}=L_{x}^{-1/2}\frac{2h^{1/2}}{1+k^{2}h^{2}}$ at large
$L_{x}$ and so
\begin{equation}
\Sigma=\frac{d^{2}}{2\pi L_{x}}\sum_{k}\frac{k^{2}}{\left(1+k^{2}h^{2}\right)^{2}}G_{0}(kR,ka).
\end{equation}
Transforming the summation into integration $\sum_{k}\rightarrow2\frac{L_{x}}{2\pi}\int_{0}^{\infty}dk$
we obtain 
\begin{equation}
\Sigma=\frac{d^{2}}{2\pi^{2}}\int_{0}^{\infty}dk\,\frac{k^{2}}{\left(1+k^{2}h^{2}\right)^{2}}G_{0}(kR,ka),\label{eq:self_gauss_CNT}
\end{equation}
This integral can be evaluated exactly in the case of point dipole
($h,R\ll a$) in cylindrical cavity with low dielectric constant,
$\epsilon_{out}\approx\epsilon_{in}$. In this case, Eq. (\ref{eq:kernel})
becomes
\begin{equation}
G_{m}(x,y)=\frac{4\pi(\epsilon_{out}-\epsilon_{in})}{\epsilon_{in}^{2}}yK_{m}(y)K_{m}'(y)\delta_{m,0}
\end{equation}
and so self-energy (\ref{eq:self_gauss_CNT}) reduces to
\begin{align}
\Sigma&=\frac{2(\epsilon_{out}-\epsilon_{in})}{\pi\epsilon_{in}^{2}}\frac{d^{2}}{a^{3}}\int_{0}^{\infty}dy\,y^{3}K_{0}(y)K_{0}'(y)\nonumber\\
&=-\frac{3\pi^{2}}{32}\frac{(\epsilon_{out}-\epsilon_{in})}{\epsilon_{in}^{2}}\frac{d^{2}}{a^{3}}.\label{eq:self_cyl_point}
\end{align}

\subsection{Gaussian envelope}

Another option for the exciton envelope wavefunction is Gaussian,
Eq. (\ref{eq:gauss_envelope}). This Gaussian envelope is chosen so
that the normalization factor $f=\frac{d}{2h^{1/2}L_{x}}$ is the
same, and, therefore, Eq. (\ref{eq:self_Bk}) applies. Comparing Eqs.
(\ref{eq:Psi_Fourier_ser}) and (\ref{eq:gauss_env_Fourier}) one
has $B_{k}=L_{x}^{-1/2}2h^{1/2}e^{-\frac{2q^{2}h^{2}}{\pi}}$ at large
$L_{x}$ and so the self-energy becomes
\begin{equation}
\Sigma=\frac{d^{2}}{2\pi^{2}}\int_{0}^{\infty}dk\,k^{2}e^{-\frac{4q^{2}h^{2}}{\pi}}G_{0}(kR,ka).\label{eq:self_gauss_cyl}
\end{equation}
Clearly, this expression converged to Eq. (\ref{eq:self_cyl_point}) when $h,R\ll a$ and the low dielectric contrast is small, since in this case the result does not depend on the envelope. Another limiting case is when $h$ is
large, so that integration is effectively limited to small momenta.
The Poisson kernel becomes
\begin{equation}
G_{0}(kR,ka)=4\pi\frac{\epsilon_{out}-\epsilon_{in}}{\epsilon_{out}\epsilon_{in}}\left(\ln ka/2+\gamma\right).
\end{equation}
Integral evaluates exactly, producing
\begin{equation}
\Sigma=-\frac{\pi}{32}\frac{\epsilon_{out}-\epsilon_{in}}{\epsilon_{out}\epsilon_{in}}\frac{d^{2}}{h^{3}}\left(\ln\frac{64h^{2}}{\pi a^{2}}-\gamma-2\right).\label{eq:self_gauss_cyl_large_h}
\end{equation}
If $\epsilon_{out}''\ll\epsilon_{out}'$, the the EVET decay rate
can be written as
\begin{equation}
\Gamma=-2\hbar^{-1}\Sigma''=\hbar^{-1}\frac{\pi}{16}\frac{\epsilon_{out}''}{\epsilon_{out}'^{2}}\frac{d^{2}}{h^{3}}\left(\ln\frac{64h^{2}}{\pi a^{2}}-\gamma-2\right).\label{eq:Gamma_gauss_cyl_large_h}
\end{equation}

\subsection{CNT EVET at realistic conditions}

\begin{figure}
\includegraphics[width=3.2in]{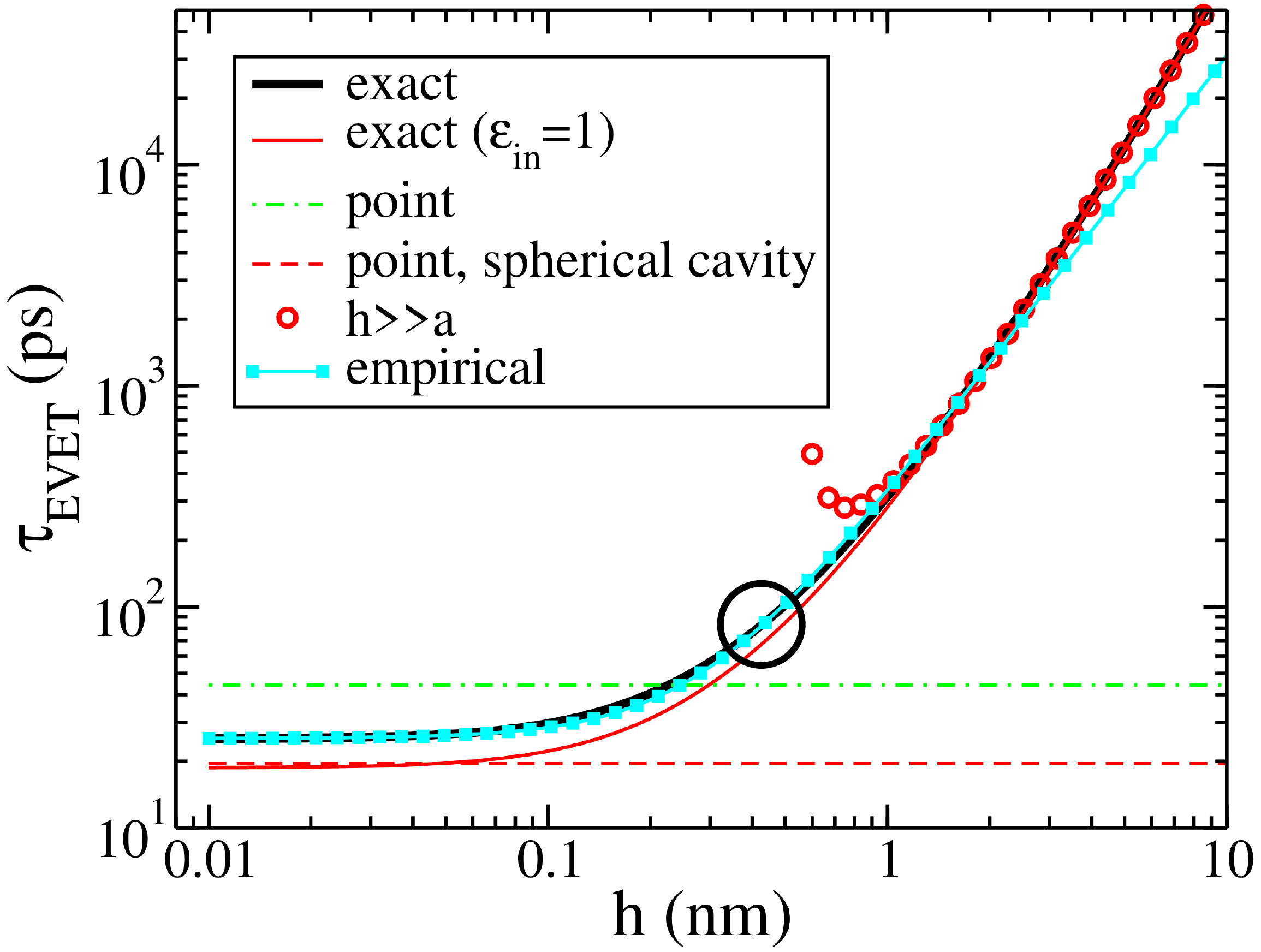}\caption{\label{fig:CNT_EVET}EVET decay time, $\tau_{{\rm EVET}}=1/\Gamma_{{\rm EVET}}$,
as a function of exciton localization length $h$. All the results
except for the thin red line are plotted for $\epsilon_{in}=1.75$.}
\end{figure}

In this subsection, we will evaluate the EVET rate corresponding to
a localized exciton in a functionalized CNT. The detailed description
of the system is given in our previous work \cite{He-2018-8060}.
Briefly, a chiral (6,5) CNT is assumed to be of radius $R=0.38\,{\rm nm}$
\cite{Saito-2000-2981,Alrawashdeh-2018-30520}. Distance between
CNT surface and the first water solvation shell (for CNTs of similar
diameters, but not specifically for (6,5) chirality) is estimated
to be $a-R\approx0.3\,{\rm nm}$ \cite{Vijayaraghavan-2014-268,Homma-2013-157402},
so we estimate the cavity radius as $a=0.68\,{\rm nm}$. CNT is functionalized
with 4-methoxybenzene, resulting in the strong localized exciton transition
at $\lambda\approx1160\,{\rm nm}$ with the transition dipole moment
of $d=38\,D$ \cite{He-2018-8060}. The dielectric constant of water
at the frequency corresponding to this transition is taken to be $\epsilon_{out}=1.75+i2.7\times10^{-5}$
\cite{He-2018-8060}. Exciton localization size in axial direction
is typically significantly larger than the CNT radius \cite{He-2017-10785,Gifford-2018-1828}.
Under these conditions, contributions of the inside of the cavity
to the overall dielectric screening is not expected to be too large and can be approximated
by a single effective dielectric constant \cite{Perebeinos-2004-257402}.
We assume two different choices for this effective dielectric constant:
(i) $\epsilon_{in}=\epsilon_{out}'=1.75$ - as if the cavity is filled
with water with no absorption, and (ii) $\epsilon_{in}=1$ - ``vacuum''
cavity. The EVET decay time, $\tau_{{\rm EVET}}=1/\Gamma_{{\rm EVET}}$,
for the just overviewed parameters is plotted in Fig. \ref{fig:CNT_EVET}
as a function of exciton localization length $h$, assuming Gaussian
exciton envelope wavefunction.
Specifically, the results of the numerical integration of Eq. (\ref{eq:self_gauss_cyl})
for the first and second choices of $\epsilon_{in}$ are plotted by
thick black and thin red lines, respectively. The dielectric screening
from inside the cavity becomes expectedly insignificant when $h\gg a$, and so
these two lines are seen to coincide at large $h$. This can be further
seen from the asymptotic expression for the EVET decay rate at large
$h$, Eq. (\ref{eq:Gamma_gauss_cyl_large_h}), that does not depend
on $\epsilon_{in}$ at all. This large-$h$ asymptotic dependence
is plotted by red circles and is seen to agree with the exact results
when $h\gtrsim 1\,{\rm nm}$. For a point dipole ($R=0$, $h=0$),
the EVET decay time for $\epsilon_{in}=1.75$ is shown by the green
dashed-dotted line. Red dashed line is the result of evaluation of
the EVET time for exactly the same parameters, but assuming a spherical
cavity of radius $a$, Eq. (\ref{eq:self_energy_spherical}). Finally,
an empirical fit to the exact numerical result for $\epsilon_{in}=1.75$ at
not too large $h$ is given by
\begin{equation}
\tau_{EVET}=1.3\tau_{EVET}^{(sphere)}\left[1+5.7(h/a)^{2}\right],
\end{equation}
where $\tau_{EVET}^{(sphere)}$ is the EVET decay time for a point
dipole in a spherical cavity (dashed red line). This empirical fit
is plotted as cyan line in Fig. \ref{fig:CNT_EVET}. Parameters $1.3$
and $5.7$ were not obtained from any rigorous fitting procedure,
but rather by manually adjusting them until the satisfactory eyeball
agreement between the cyan and black lines at $h\lesssim 3\,{\rm nm}$ is
reached. These parameters are in principle functions of $R/a$. In
particular, at $R\rightarrow0$, the prefactor of $1.3$ has to be
substituted with the one that superimposes cyan and dashed-dotted
green lines at $h\rightarrow0$. 

To obtain the exciton localization length $h$ one needs to fit the
exciton transition charge density, obtained for example from TD-DFT
calculations, with Eq. (\ref{eq:exp_envelope}) or Eq. (\ref{eq:gauss_envelope}).
Using exactly this approach we obtained $h\approx0.54\,{\rm nm}$
and $h\approx0.43\,{\rm nm}$ for the exponential and Gaussian envelopes,
respectively \cite{Gifford-2019-unpublished}. This produces the
EVET decay times of 120 and 85 ps, respectively, for $\epsilon_{in}=1.75$.
In the case of $\epsilon_{in}=1$, the EVET decay rate drops from 85
to 70 ps. The results for $h\approx0.4\,{\rm nm}$ for the Gaussian
envelope are shown by a large black circle in Fig. \ref{fig:CNT_EVET}.
Experimentally, the EVET decay rate in water was estimated to be $\approx200\,{\rm ps}$
in Ref. \cite{He-2018-8060}. Considering the level of approximations
in this work, the obtained theoretical result ($\tau_{{\rm EVET}}\sim100\,{\rm ps}$)
is in good agreement with experiment.

\section{Conclusion\label{sec:Conclusion}}

In this work we developed a general theory of EVET-mediated relaxation
of localized excitons in functionalized CNTs. General expressions
for the EVET rate, as well as estimates of the EVET exciton relaxation
times at realistic conditions were obtained in Sec. \ref{sec:EVET-CNT}.
The specific result, $\tau_{EVET}\sim100\,{\rm ps}$, is in good agreement
with previous experimental results \cite{He-2018-8060}, considering
the level of approximations in this work. The most straightforward
way to discuss the level of approximations and the possible approaches
to more accurate computations is to consider Eq. (\ref{eq:dec_rate}).
In this very general expression, the EVET rate depends on exciton
transition charge density and the Green's function of the electrostatic
Poisson equation in the cavity corresponding to the CNT. The former,
in the form of the transition dipole moment density (\ref{eq:P_Phi_ex}),
was assumed to be a smooth function slowly varying on the scale of
the unit cell, and either of exponential or Gaussian envelope. Another,
more accurate, approach would be to extract the transition charge
density directly from e.g., time-dependent DFT calculations \cite{He-2017-10785,Gifford-2018-1828}.

We found the Green's function of the Poisson equation assuming perfectly cylindrical
cavity. This approximation breaks down near the functional group attached
to the CNT. The presence of the functional group should somewhat increase
the effective solvation radius as solvent molecules cannot be present
in the volume occupied by the functional groups. Again, the remedy
can come from electronic structure theory calculations, where the
Poisson equation is solved routinely for arbitrary shaped cavities
within the polarizable continuum model framework \cite{Klamt-1993-799,Mennucci-2012-386}.
An additional complication is the dielectric screening contributed
by the CNT itself, which should in principle be incorporated in a
non-local and possibly even non-static manner. 

\appendix

\section{Transition Charge Density for CNT exciton\label{sec:CNT_trans_dens}}

Within the 2-band $k\cdot p$ Hamiltonian method, lowest conduction
and highest valence band single-particle CNT wavefunctions are ($qR\ll1$)
\cite{Ando-2005-777}
\begin{gather}
\Psi_{c,k_{c}}(x,\xi)=\frac{1}{L_{x}^{1/2}L_{\xi}^{1/2}}\left(u_{c}+\frac{\hbar k_{c}}{\sqrt{2m_{*}\Delta}}u_{v}\right)e^{ik_{c}x}e^{iQ_{x}x+iQ_{\xi}\xi},\label{eq:Psi_sp_CNT}\\
\Psi_{v,k_{v}}(x,\xi)=\frac{-i}{L_{x}^{1/2}L_{\xi}^{1/2}}\left(u_{v}-\frac{\hbar k_{v}}{\sqrt{2m_{*}\Delta}}u_{c}\right)e^{ik_{v}x}e^{iQ_{x}x+iQ_{\xi}\xi},\nonumber 
\end{gather}
with corresponding energies $E_{q}=\pm\left(\frac{\Delta}{2}+\frac{\hbar^{2}q^{2}}{2m_{*}}\right)$.
Single-particle effective mass and bandgap energy are denoted by $m_{*}$
and $\Delta$, respectively. Bloch function spinors are $u_{c}=[1,0]^{T}$,
$u_{v}=[0,1]^{T}$. Constant wavenumbers $Q_{x}$, $Q_{\xi}$ are
determined by (i) the CNT chirality, (ii) choice of valley ($K$ or
$K'$). In what follows, we assume only optically allowed ``vertical''
exciton transitions so that $Q_{x,\xi}$ does not change upon transition
from the valence to conduction band. Spin indices are omitted since
the final result does not depend on them. Seemingly unnecessary factor
of $-i$ in the valence-band wavefunction in Eq. (\ref{eq:Psi_sp_CNT})
is added so that the resulting transition density is real (see below).

Within the multi-band envelope function method, the charge density
operator is proportional to the unit matrix \cite{Velizhanin-2016-165}.
Transition charge density, Eq. (\ref{eq:tr_charge_dens}), is then
straightforwardly generalized as $\rho(x,\xi)=-e\sum_{k_{c},k_{v}}B_{k_{c}k_{v}}\Psi_{c,k_{c}}^{\dagger}(x,\xi)\sigma_{0}\Psi_{v,k_{v}}(x,\xi)$.
Here, $\sigma_{0}$ is $2\times2$ unit matrix, and the Hermitian
conjugation of $\Psi_{c,k_{c}}$ results not only in complex conjugation,
but also in transposition of Bloch function spinors, that is e.g.,
$u_{c}^{\dagger}=[1,0]$ - a row vector. More explicitly, the exciton
transition density is
\begin{equation}
\rho(x,\xi)=e\sum_{k_{c},k_{v}}B_{k_{c}k_{v}}\frac{i}{L_{x}L_{\xi}}e^{-i(k_{c}-k_{v})x}\frac{\hbar(k_{c}-k_{v})}{\sqrt{2m_{*}\Delta}},\label{eq:rho_exc_cv}
\end{equation}
where coefficients $B_{k_{c}k_{v}}$ can generally be obtained by
solving Bethe-Salpeter equation. We, however, first consider a freely
propagating exciton of momentum $k$ and assume a simple effective
mass approximation where the electron and hole relative motion is
hydrogen atom-like and does not depend on the total exciton momentum.
Under these conditions, the electron-hole wavefunction for such exciton
is $\Psi_{k}(x_{e},x_{h})=L_{x}^{-1/2}L_{\xi}^{-1}\phi_{eh}(x_{e}-x_{h})e^{ik(x_{e}+x_{h})/2}$,
where $\phi_{eh}(x_{e}-x_{h})$ encodes the electron and hole axial
relative motion in the center-of-mass reference frame, and $e^{ik(x_{e}+x_{h})/2}$
represents the motion of the exciton as the whole. This wavefunction
can be expanded into single-particle plane waves as
\begin{equation}
\Psi_{k}(x_{e},x_{h})=\sum_{k_{e},k_{h}}B'_{k_{e}k_{h}}\frac{e^{ik_{e}x_{e}}}{L_{x}^{1/2}L_{\xi}^{1/2}}\frac{e^{ik_{h}x_{h}}}{L_{x}^{1/2}L_{\xi}^{1/2}},\label{eq:Psi_exc_eh}
\end{equation}
so the expansion coefficients are
\begin{align}
B'_{k_{e}k_{h}}&=\frac{1}{L_{x}L_{\xi}}\int dx_{e}d\xi_{e}\int dx_{h}d\xi_{h}\,\Psi_{k}(x_{e},x_{h})e^{-ik_{e}x_{e}}e^{-ik_{h}x_{h}}\nonumber\\
&=\delta_{k,k_{e}+k_{h}}B'_{k_{e}-k/2},
\end{align}
where
\begin{equation}
B'_{q}=\frac{1}{L_{x}^{1/2}}\int dx\,\phi_{eh}(x)e^{-iqx}.
\end{equation}
Expansion coefficients of the exciton wavefunction into single-particle
states in Eqs. (\ref{eq:rho_exc_cv}) and (\ref{eq:Psi_exc_eh}) are
related as $B_{k_{c}k_{v}}=B'_{k_{c},-k_{v}}$, and therefore the
transition charge density for the exciton of momentum $k$ becomes
\begin{equation}
\rho_{k}(x,\xi)=e\phi_{eh}(0)\frac{i}{L_{x}^{1/2}L_{\xi}}e^{-ikx}\frac{\hbar k}{\sqrt{2m_{*}\Delta}}\label{eq:rho_k_phi0}
\end{equation}
Now, if we consider a localized (by a trap) exciton where the localization
is shallow enough to not affect the exciton internal structure and
so $\phi_{eh}$ is independent of $k$ in Eq. (\ref{eq:rho_k_phi0}),
the transition charge density for this localized exciton is
\begin{equation}
\rho(x,\xi)=\sum_{k}A_{k}^{*}\rho_{k}(x,\xi),
\end{equation}
where 
\begin{equation}
A_{k}=\frac{1}{L_{x}^{1/2}}\int dx\,\Phi(x)e^{-ikx}
\end{equation}
 and $\Phi(x)$ is an envelope function of the localized exciton.
The function $\Phi(x)$ is real and so $A_{k}=A_{-k}^{*}$. The transition
charge density is
\begin{equation}
\rho(x,\xi)=-e\frac{1}{L_{x}^{1/2}L_{\xi}}\phi_{eh}(0)\sum_{k}A_{k}^{*}e^{-ikx}\frac{-i\hbar k}{\sqrt{2m_{*}\Delta}}.
\end{equation}
The summation can be represented via the first derivation of $\Phi$
with respect to $x$
\begin{equation}
\rho(x,\xi)=-e\frac{1}{L_{\xi}}\frac{\hbar}{\sqrt{2m_{*}\Delta}}\phi_{eh}(0)\frac{\Phi(x)}{dx}.
\end{equation}
Transition dipole moment density is defined as $\rho=-\nabla\cdot{\bf P}$,
and so
\begin{equation}
{\bf P}(x,\xi)=\frac{1}{L_{\xi}}\frac{\hbar e}{\sqrt{2m_{*}\Delta}}\phi_{eh}(0)\Phi(x){\bf e}_{x},\label{eq:P_prop_Phi}
\end{equation}
where ${\bf e}_{x}$ is a unit vector parallel to the $x$-axis. This
very simple result means that the transition dipole moment density
does not depend on circumferential coordinate $\xi$ and linearly
proportional to $\Phi(x)$ - the exciton envelope wavefunction. That
${\bf P}(x,\xi)$ is slowly varying on the scale of the unit cell
here can be traced back to the multi-band envelope function $k\cdot p$
approximation, where the transition charge density operator is proportional
to the unit matrix and, therefore, the specific local structure of
the transition charge density on the scale of single unit cell is
averaged over \cite{Velizhanin-2016-165}.


%

\end{document}